\documentstyle[11pt]{article}
\begin{document}
\begin{titlepage}

\hspace{11cm}ADP-95-16/T177

\vspace{1cm}

\centerline{\Large \bf POSSIBLE EFFECTS OF QUARK AND GLUON CONDENSATES}
\centerline{\Large \bf IN HEAVY QUARKONIUM SPECTRA\footnote{This work was
partly supported by the National
Natural Science Foundation of China (NSFC) and the Australian Research
Council}}
\vspace{1cm}
\centerline{ Y.B. Ding$^{a,b,c}$,
X.H. Guo$^{c,e}$, X.Q. Li$^{a,d,c}$, P.N. Shen$^{a,c,e,f}$}
\vspace{1cm}
{\small
{
\flushleft{\bf  $~~~a.$ China Center of Advanced Science and Technology
 (World Laboratory),}
\flushleft{\bf  $~~~~~~$P.O.Box 8730, Beijing 100080, China}
\vspace{8pt}
\flushleft{\bf  $~~~b.$ Graduate School, Academia Sinica, P.O. Box 3908,
 Beijing 100039, China}
\vspace{8pt}
\flushleft{\bf  $~~~c.$ Institute of Theoretical Physics, Academia Sinica,
 P.O.Box 2735,}
\flushleft{\bf  $~~~~~~$Beijing 100080, China}
\vspace{8pt}
\flushleft{\bf  $~~~$d. Department of Physics, Nankai University, Tianjin 
300071, China}
\vspace{8pt}
\flushleft{\bf  $~~~$e. Institute of High Energy Physics, Academia Sinica,
 P.O.Box 918(4),}
\flushleft{\bf  $~~~~~~$Beijing 100039, China}
\vspace{8pt}
\flushleft{\bf  $~~~$f. Department of Physics and Mathematical Physics, the
 University of Adelaide,}
\flushleft{\bf  $~~~~~~$GPO Box 498, Adelaide, South Australia 5001}
}}

\vspace{2cm}

\centerline{\bf Abstract}

\noindent
The Cornell potential with the best fitted parameters
$\alpha_{s}$ and $\kappa$ are modified by adding terms derived from 
non-perturbative QCD, which are characterized by a series
of non-vanishing vacuum condensates of quarks and gluons. In terms of this
potential,
we study the system of 
heavy quarkonia. The results show that the correction caused by the
additional terms reduces the deviation between the data and the
values calculated with the pure Cornell potential 
and improves the splittings of energy levels. The achievements indicate 
that the non-perturbative effects induced by vacuum condensates
play an important role for the correction to $1/q^2$, which in general
was phenomenologically
put in by hand. This result would be helpful for understanding
non-perturbative QCD along a parallel direction to the QCD sum rules.

\end{titlepage}

\baselineskip 18pt

\noindent {\bf I. Introduction}

\vspace{0.5cm}

The success of the potential model
in explaining hadronic spectra and
hadronic properties has been remarkabe. Especially, for the
$J/{\Psi}$ and ${\Upsilon}$ families, various potential
forms [1], in which both the Coulomb and confinement
potentials are employed, can give results which are 
reasonably consistent with the data at the charm and bottom energy
scales within a certain error. In general, it is believed that confinement
comes from the non-perturbative effect of QCD, but
unfortunately, so far, that is an unsolved problem.

Along another line, years ago, Shifman et al. \cite{svz} proposed
to study hadronic
properties in terms of the QCD sum rules where a few non-vanishing vacuum
condensates of quarks and gluons 
$m_{q} < \psi_{q}\bar{\psi}_{q} >$,
${\alpha_{s}\over\pi} <G^{a}_{\mu\nu}
G^{a\mu\nu}>$ etc. describe the nonperturbative effect. They started from
short distance, where the quark-gluon dynamics is essentially
perturbative, and extrapolated the dynamics to larger distance
by introducing non-perturbative effects step 
by step [3]. The applications of the theory extrapolated in studying hadronic
properties such as the spectrum, decay width and hadronic matrix elements
etc., indicate that one can trust the validity of this approach.

Inspired by these successes, we have been trying to
introduce the non-perturbative QCD effects,
characterized by non-vanishing quark and gluon vacuum
condensates, into the traditional
potential model [4], so that a deeper understanding of
the hadronic structure and the underlying mechanisms which determine how
quarks are bound into hadrons can be obtained.
In our derivation of correction terms, the vacuum condensates were employed
to modify the free gluon propagator. As a consequence, the quark-quark
potential and the spectrum of heavy quarkonium are affected. Studying
these possible effects is the main purpose of this paper. Meanwhile,
we have noticed that beside the modification of the gluon propagator,
characterized by the vaccum condensates, the closed-loop correction can also
contribute a comparable effect, because these two kinds of corrections are
in the same order of $\alpha_{s}$. This has been shown by Gupta et al [5,6],
Fulcher [7] and Pantaleone et al [8]. In this investigation, as a
preliminary study, we tempararily treat the condensate effect alone and
will collect the closed-loop correction in our later paper.

As to the condensate correction, there are two different results, ours [4]
and Larsson's [9]. The effect of the difference may show up in the
spin splittings of heavy quarkonia. Thus our additional effort
in this paper will be donated to search the possible effects of
these two approaches. We hope this investigation may help us to understand
the condensate correction deeply.

As mentioned above, the non-vanishing vacuum condensates can only be
used to describe an extrapolation from the short distance to the
medium range [3], therefore,
the longer distance effect at the energy scale $\leq\Lambda_{QCD}$
cannot be included in the scenario. In other
words, this scenario is only valid within the intermediate range if
only a finite number of vacuum condensates are kept,
and for the bound states,
the potential term responsible for the confinement should come from the
larger distance$1/\geq\Lambda_{QCD}$. Therefore, we are attempting
to introduce a reasonable picture where the main contribution to the
confinement is caused by the interaction at $\leq\Lambda_{QCD}$,
where the physical picture
is not clear yet. Since this part is not derivable at the present
stage, we keep a phenomenological confinement form, generally
the linear $\kappa r$ term which is the main part of the confinement
potential and universal to all of the heavy flavors.    
Then we perturbatively introduce the corrections induced by the non-trivial
vacuum condensates into our framework. It would make
observable contributions to some hadronic
properties of $J/\Psi$ and $\Upsilon$ families, for example the spin
splitting between 1$^{3}S_{1}$ and 1$^{1}S_{0}$ could be one of the 
sensitive quantities for the correction.
Moreover, we would claim that there is no double-counting between
our derived correction and the contribution from the larger
distance $1/\Lambda_{QCD}$.

It  should be emphasized that the amplitude of the wave function at the
origin, which is essential to the spin splitting, depends on the potential
model adopted. For instance, Richardson potential [10] wave function at
the origin is almost the half of Cornell's. In this investigation, we
start with the Cornell potential, which has the simplest form  among
existant potential models, as the zeroth order approximation and then
add in the condensate corrections in a perturbative way. Therefore, our
wave function at the origin is close to that of the Cornell potential,
and the difference between ours and Cornell's starts to show up only
in the approximation higher than the zeroth order. In our later work,
we will study these differences and effects caused by adopting different
model wave function as the zeroth order wave function.

In the next section, the potential corrections derived with
vacuum condensates are briefly reviewed and  different formulae are analyzed. 
In section III, the numerical results are presented and compared 
with the data. Finally, our results are discussed and conclusions are drawn.\\

\noindent {\bf II. Our Model}

\vspace{0.5cm}

As SVZ \cite{svz} suggested, there are non-zero vacuum expectation values of 
the quark ($<\bar{\psi}_{q}\psi_{q}>$) and
gluon ($\frac{\alpha_{s}}{\pi}<GG>$)
fields, so the propagator of the gluon should include
the effects of these condensates. In fact, in the propagator, all the
terms associated with vacuum condensates are proportional to
$\alpha_{s}^{n}$ $(n\geq 1)$. On the other hand, the contributions of 
higher order perturbative QCD
corrections to the potential were discussed in Ref.[5]. In comparison
with such corrections, the condensate terms do not suffer from the loop
suppression. SVZ noticed that fact and then suggested that
the condensates could be considered as a larger
contribution. Therefore, at the order of $\alpha_{s}^{2}$, we only include
the condensates, but not the perturbative correction.

There have been various ways to modify the potential. Except the higher order
perturbative QCD correction, all of them
attempt to include some non-perturbative effects to the potential, because
it is sure that such effects must be taken into account.
Richardson et al proposed a form \cite{ric}
$$\tilde U(q^2)=-{4\over 3}\frac{12\pi}{(33-2N_f)}{1\over q^2}{1\over
ln(1+{q^2\over\Lambda^2})}$$
where $q^2$ is the momentum of the gluon exchanged between quarks and $N_f$
represents the flavor number, while the linear confinement still remains
unchanged as $\kappa r$. It gives an effective correction which indeed involves
the non-perturbative effects. Besides, Fulcher \cite{ful} gave
$$V(r)=Ar-{8\pi\over (33-2N_f)}{1\over r}f(\Lambda r)$$
where $f(\Lambda r)$ has a very complicated integration 
form, which can be found in ref.\cite{ful}. Recently, Gupta et al. considered
not only the higher order perturbative radiative corrections, but also a
more complicated non-perturbative term \cite{gup}.
In all these works, the corrections related to non-perturbative effects are
phenomenologically put in according to the observation  or hint from
lattice gauge results. Some very good results which coincide with
data within an error of a few MeV were reported \cite{gup}.
We will discuss this problem in some detail
in the last section.

On the other hand, in this work we are trying to understand 
such corrections in terms of
some well-established theories which can handle the non-perturbative QCD in
a more natural way.

Within the QCD scenario, where the non-vanishing vacuum condensates of
quarks and gluons characterize the non-perturbative effects,
the modified gluon propagator in momentum space can be written as [4]
\begin{eqnarray}
G_{\mu\nu}=\frac{-i}{q^{2}}(g_{\mu\nu}-\frac{q_{\mu}q_{\nu}}{q^{2}})F(q^{2})
\end{eqnarray}
where
\begin{eqnarray}
F(q^{2}) = 1+\frac{1}{3} g^{2}_{s} \sum_{\beta=u,d,s} \frac{m_{\beta}
<\psi_{\beta}\bar{\psi_{\beta}}>}{q^{2}(q^{2}-m^{2}_{\beta})} ~+~
\frac{9}{32}g^{2}_{s} <G^{2}> \frac{1}{q^{4}}.
\end{eqnarray}
In this expression we only keep the lowest dimensional condensates
$<\psi_{q}\bar{\psi}_{q}>$ and $<GG>$. We
derived this expression in the standard way [4], in which the normal
product operators such as $\psi \bar{\psi}(0)$ and $G^{2}(0)$ have
non-vanishing matrix elements in the physical vacuum, --- i.e.,
$<\psi\bar{\psi}>$ and $<G^{2}>$ are left as parameters to describe
non-perturbative effects, and their values have already been determined in
literature \cite{svz}. 

In another way, by  comparing the $(2n+1)$-point
Green's function ($n$ is the number of external legs of $\psi$ or $A^{\mu}$)
and the $n$-point Green's function with the insertion of the operators
$\psi\bar{\psi}(0)$ or $G^{2}(0)$ (these Green's functions are with
respect to the physical vacuum), Larsson achieved \cite{lar}
\begin{eqnarray}
D_{\mu \nu}~=~ [ 1~-~\sum_{\beta} \frac{g_{s}^{2} m_{\beta}
<\psi_{\beta}\bar{\psi_{\beta}}>}{q^{2} (q^{2}+m^{2}_{\beta})}~+~
\frac{5g_{s}^{2} < G^{2}>}
{288q^{4}} ]^{-1} \frac{(-i)}{q^{2}} (\delta_{\mu \nu} - \frac{q_{\mu}q_{\nu}}
{q^{2}})
\end{eqnarray}
where the propagator was derived in the Euclidean space. Expanding the
first piece with respect to $\alpha_{s}$, one notices that not only the
coefficients at the lowest order of $\alpha_{s}$ are 
different from those in Eq.(2),
but also the sign of the coefficient associated with the gluon condensate is
different from that in Eq.(2). These differences  arise from different 
approaches used to deal with the
non-perturbative effects and perhaps due to the the improper usage of the
fixed point gauge \cite{svz} which violates translational invariance.
\cite{henz}

In this work, we also determine phenomenologically whether Eq.(2) or Eq.(3) 
is more consistent with the data.

We write down a proper scattering amplitude
between two quarks as
\begin{eqnarray}
M=(-ig_{s})^{2}\bar{u}_{q}(p_{1})\gamma_{\mu}\frac{\lambda^{a}}{2}
u_{q}(p_{1}')D^{\mu\nu}(q^{2})\bar{u}_{q'}(p_{2})\gamma_{\nu}
\frac{\lambda^{a}}{2}u_{q'}(p_{2}')
\end{eqnarray}
with
$$p_{1}-p_{1}'=p_{2}'-p_{2}=q.$$
and  carry out the Breit-Fermi expansion with the spinors $u_{q}(p_{i})$
being the solutions of free quarks. In deriving an effective potential,
the  spontaneous approximation, $q_{0}=0$, has been taken.
It should be mentioned that this approximation is a traditional
treatment in literatures although it is not quite valid in the potential
derivation. Moreover, in ref.\cite{lan},
the term $q_{\mu}q_{\nu}$ was kept as
$q_{i}q_{j}$ $(q_{0}=0)$. If the CVC theorem is respected, i.e.
$q^{\mu} \bar{u}_{q}(p_{1}) \gamma_{\mu} u_{q}(p_{1}')=0$,
the term $q_{\mu}q_{\nu}/q^{2}$ vanishes, and then the derived potential
will have a small difference with that without CVC theorem, 
even though it is not very extravagantly apart.
Then one can apply the three-dimensional Fourier transformation
to convert the propagator in momentum space to coordinate space
and derive an effective potential between quarks.

The potential derived in the way used in ref.[4], in which leading order
non-perturbative QCD effects are considered,
together with the phenomenological
linear confinement can be written in the following form:
$$
V(r)=-\frac{4\alpha_{s}}{3r}+\kappa r+V^{corr}_{1}(r)+V^{corr}_{2}(r)
+V^{corr}_{3}(r),
$$
where $V^{corr}_{1}$ is the correction from the non-trivial physical
vacuum condensates, while $V^{corr}_{2}$ and $V^{corr}_{3}$ are the 
Breit-Fermi corrections to the Coulomb and linear confinement terms, 
respectively. Due to the lengthy expressions for these potentials, we do
not present them here. The explicit forms of these potentials can be found
in Appendix.

It is noted that:

(1). In the preceding literature, for instance ref.[14],
a comparison with the experimental data shows that the
potential $\kappa r-\frac{4\alpha_{s}}{3r}$,
with universal values of $\kappa$ and $\alpha_{s}$ is applicable to both
$J/\Psi$ and $\Upsilon$ in  solving the Schr\"{o}dinger
equation. Hence, this confining potential should be independent 
of the quark mass
$m_{c}$ or $m_{b}$. In other words, it corresponds to the potential in the
limit $m_{Q}\rightarrow\infty$ $(m_{Q}\gg\Lambda_{QCD})$. It is universal
and exists in all $Q\overline{Q}$ systems. On the other hand, all correction
terms associated with vacuum condensates are of the order of $1/m^{2}_{Q}$,
which can be directly read from the Feynman diagrams.
Therefore, they appear as mass-dependent corrections to 
$\kappa r-\frac{4\alpha_{s}}{3r}$. Consequently, there is no
double-counting involved in $\kappa r-\frac{4\alpha_{s}}{3r}$. Moreover,
the Coulomb term ${-4\alpha_s\over 3r}$ comes from one-gluon-exchange and
represents a short-distance effect where perturbative QCD works
perfectly well. Therefore the correction induced by the condensates which
manifest the non-perturbative effects does not overlap with the pure
Coulomb part either. The $\kappa r$ term 
may be understood in the following picture. When
$m_{Q}$ is very large, one can define the total momentum $k$ of the quark
$Q$ as $P_{Q}=m_{Q}v+\delta$ where $Q$ is almost on the mass shell,
$v$ is the four-velocity of $Q$ and
$\delta$ is the so-called residual momentum of the order of
$\Lambda_{QCD}$ \cite{geo}.
>From Eq.(4), one finds that in the large $m_Q$ limit the emitted gluons are
soft. This leads to the conclusion that the confinement term, $\kappa r$,
plays a role at the
energy scale $\Lambda_{QCD}$. In this picture, $\kappa r$ is 
independent of $m_{Q}$.

(2). Since all the correction terms associated with the vacuum
condensates are proportional to $1/m^{2}_{Q}$, under this meaning,
they are of the same order as the
relativistic corrections in the Breit-Fermi expansion.

(3). Since the higher order terms are omitted, the derived potential
is not appropriate in dealing with higher resonances.

(4). Our numerical results indicate that beyond the 2S state the calculated
values would deviate from the data more and more, but for the 1S, 1P and
2S states, these values indeed make sense (see below).\\

\noindent{\bf III. Numerical Results}

\vspace{0.5cm}

In this work, to elucidate the significance
of the correction, we present a few very typical quantities which are
calculated in the framework of QCD.

The Cornell potential $-4\alpha_{s}/3r~+~\kappa r$
with corresponding parameters $\alpha_{s}$ and $\kappa$ \cite{lic}, which gave
the best fit to the $J/\Psi$ and $\Upsilon$ family data, is adopted as a
basic condition, and the values of vacuum condensates are taken 
from ref.\cite{svz}. Thus there is no free parameters at all in 
the derived expressions (2) and (3).

To our understanding, the term
$\kappa r$  is universal to $J/\Psi$ and $\Upsilon$
and dominates the confinement part. This is consistent with the consideration
in the literature
which deals with non-perturbative corrections. Then one can consider
the additional part from condensates
as a $1/m^{2}_{Q}$ correction to the potential. It is noticed that,
as discussed in most of the preceding literatures, one always assumed that
$\kappa r$ was
caused by a scalar exchange. As a consequence, it would not induce a spin
splitting. On the other hand, the only term responsible
for the spin splitting is the Coulomb term which contributes as a vector
potential. It should be mentioned that
that assumption was based on phenomenological requirements --- i.e. for
the Cornell potential, if $\kappa r$ was induced by scalar 
exchange, a  better fit to the spin splitting data could be obtained.
Now, the new correction  induced by the condensates contributes
to the spin splitting, so the whole picture changes. In particular,
this modification demands that $\kappa r$ comes not only from scalar
exchange, but also from vector exchange. Thus, we can write
\begin{equation}
\label{kr}
 \kappa r=\beta \kappa r+(1-\beta)\kappa r\;\;\;\;\; 0\leq \beta\leq 1
\end{equation}
where the factor $\beta$ characterizes the fraction of the confinement 
potential which comes from vector
exchange, while $(1-\beta)$ denotes that from scalar exchange. If 
$\kappa r$ is fully caused by the scalar exchange, then $\beta=0$ and 
it is the same assumption as in the preceding literature. The explicit value
of $\beta$ can be fixed by data fitting. Then the spin splitting
$\Delta=(M_{1^{3}S_{1}}-M_{1^{1}S_{0}})$ for $c\bar{c}$ and $b\bar{b}$ systems
can be calculated. The correction term
which contributes to the spin splitting can be read as
\begin{eqnarray}
-(\frac{g^{2}}{4\pi})\frac{<\lambda^{a}\lambda^{a}>}{4}
[(\frac{1}{6}\sum_{\beta} a_{\beta} \frac{1}{r}-\frac{1}{12}br
-\frac{1}{6}\sum_{\beta} a_{\beta} \frac{e^{-m_{\beta}r}}{r})
+\frac{\pi}{3}c\delta(\vec{r})
](\vec{\sigma_{1}}\cdot\vec{\sigma_{2}}),
\end{eqnarray}

\noindent
where
$$c=\frac{2}{m_{1}m_{2}} +
\sum_{\beta} \frac{A'_{\beta}}{m_{\beta}^{3}m_{1}m_{2}},
~~~a_{\beta}~=~\frac{A'_{\beta}}{m_{1}
m_{2}m_{\beta}},~~~b~=~\frac{B'}{m_{1}m_{2}},$$

$$
A'_{\beta}=\left\{ \begin{array}{ll}
                     \frac{g_{s}^{2}}{3}<\psi_{\beta}\bar{\psi_{\beta}}>
                     &\mbox{ref.[4]}\\
                     g_{s}^{2}<\psi_{\beta}\bar{\psi_{\beta}}>
                     &\mbox{ref.\cite{lar}}
                     \end{array} \right.
\hspace{0.5cm}
B'=\left\{ \begin{array}{ll}
             \frac{9}{32}g_{s}^{2}<G^{2}>   &\mbox{ref.[4]}\\
             -\frac{5}{288}g_{s}^{2}<G^{2}> &\mbox{ref.\cite{lar}}
             \end{array} \right.
$$

\noindent
and $m_{1}=m_{2}$ are the masses of the heavy quarks in quarkonium.

It is noted that, the terms with $a_{\beta}$ in refs.[4] and \cite{lar} have
the same sign, but the numerical value of this term in ref.\cite{lar}
is three times larger than that in ref.[4]. Since the contribution from this
term has the same sign (though $<\psi\bar{\psi}>$)  as that
from the term with $\delta(\vec{r})~~(<\psi\bar{\psi}>~>~0),$
it enhances the spin splitting
between $1^{1}S_{0}~(~<\vec{\sigma} \cdot \vec{\sigma}>~=~-3~)$ and
$1^{3}S_{1}~(~<\vec{\sigma} \cdot \vec{\sigma}>~=~1~)$.
However, for the terms with $b$ in refs.[4] and \cite{lar}, they have not
only different numerical values, but also opposite signs. Therefore, 
the term with $b$ given by ref.\cite{lar}
tends to increase the spin splitting, while the corresponding term
in ref.[4] reduces the spin splitting. Since
$$\frac{(5/288)<G^{2}>r^{2}}{<\psi_{\beta}\bar
\psi_{\beta}>/m_{\beta}}\Bigg\vert_{r~\sim~0.4fm}~\sim~0.016(
\frac{\pi}{\alpha_{s}})$$
is a small number, the existence of $<GG>$ does not give rise to a
more significant
influence to the spin splitting than that of $<\psi_q\bar\psi_q>$.
Anyway, a measurement of the spin splitting between
$1^{1}S_{0}$ and $1^{3}S_{1}$  tells us that the non-perturbative
effects characterized by the non-vanishing vacuum condensates indeed
play an important role, so manifest themselves in phenomenology. 

The spin splitting of $1^3P_{J}$ state  is also influenced  by the above
mentioned correction as well as the parameter $\beta$. The correction terms 
can be re-written in the following way:
\begin{eqnarray}
V^{corr}_{1}(r) &=&  
V^{cen}_{1}(r)+V^{s-o}_{1}(r)+V^{ten}_{1}(r)+\cdot\cdot\cdot \nonumber \\
V^{corr}_{2}(r) &=&  
V^{cen}_{2}(r)+V^{s-o}_{2}(r)+V^{ten}_{2}(r)+\cdot\cdot\cdot\\ 
V^{corr}_{3}(r) &=&  
V^{s-o}_{3}(r)+V^{ten}_{3}(r). \nonumber
\end{eqnarray}
with
\begin{eqnarray}
\label{vec}
V^{s-o}_{3} &=& {1\over 2m_Q^2}\left[3{V'\over r}-{S'\over r}\right] \nonumber \\
V^{ten}_{3} &=& {1\over 12 m_Q^2}\left[{V'\over r}-V^{''}\right],
\end{eqnarray}
where the superscripts $cen, s-o$ and $ten$ denotes the central, spin-orbit 
and tensor parts of the corresponding $V^{corr}_{i}(r)$, respectively, 
and $V$ and $S$ represent the vector and scalar parts of the confinement
$\kappa r$ in Eq.(5), respectively. Then the spectrum of $1^{3}P_{J}$ states
can be expressed as 
\begin{eqnarray}
M(1^3P_2) &=& \overline M+f-{2\over 5}g \nonumber \\
M(1^3p_1) &=& \overline M-f+2g \\
M(1^3p_0) &=& \overline M-2f-4g, \nonumber
\end{eqnarray}
where
\begin{eqnarray}
\label{vec}
f &=& <V^{s-o}_{1}(r)>+<V^{s-o}_{2}(r)>+<V^{s-o}_{3}(r)> \nonumber \\
g &=& <V^{ten}_{1}(r)>+<V^{ten}_{2}(r)>+<V^{ten}_{3}(r)>,
\end{eqnarray}
and $\overline M$ is the weighted average for the $1^3P_{J}$ states.

It is noticed that  in Eq. (\ref{vec}) the $V$ and $S$ 
terms are of opposite sign
and the newly achieved  correction term is opposite in sign to the
Coulomb term. Then one may easily be convinced that to keep a good fit to
data, the appearance of non-perturbative correction terms 
requires $\beta\neq 0$. In fact, Gupta et al. \cite{gup} also noticed 
that with a reasonable contribution from non-perturbative effects, 
the linear confinement $\kappa r$ must be a superposition of two 
parts as given in Eq.(5). As a result, they obtained an approximate 
$\beta-$value of 0.25 with their model potential. Alternatively, with
our modified potential, we have $\beta\approx0.5\sim 0.6$
for a better fit.

The numerical calculation is performed in the following way. The
Cornell potential which is universal to $c$ and $b$ families,
i.e., independent of $m_Q$, is considered as the dominant  part of the
potential, and the corresponding parameters $\alpha_s$ and $\kappa$
are determined before adding in the corrections. The values of $\alpha_s$
and $\kappa$ are 0.381 and 0.182$GeV^{2}$, respectively.
And then, the newly derived corrections
due to the non-vanishing vacuum condensates as well
as the Breit-Fermi corrections are treated as a perturbation adding onto 
the dominant part. The resultant values for the $c\bar{c}$ and $b\bar{b}$
systems are tabulated in Table Ia and Ib, respectively.

\vspace{0.5cm}

\newpage

\centerline{Table Ia. $c\bar{c}$ system}

\vspace{0.3cm}

\begin{footnotesize}
\begin{center}
\begin{tabular}{|c|c|c|c|c|c|c|}
\hline
 & $~~~~~~~$ & $~~~~~~$ & $V^{Cornell}+V^{corr}_{2}$ &
  $V^{Cornell}+V^{corr}_{2}$ & $~~~~~~~~$ &
    $~~~~~~~~$ \\
 & exp't.  & $V^{Cornell}=$ & $+V^{corr}_{3}+V^{corr}_{1}$ &
  $+V^{corr}_{3}+V^{corr}_{1}$ & $V^{Cornell}+V^{corr}_{2}$ &
   $V^{Cornell}+V^{corr}_{2}$ \\
 &  & ${-4\alpha_s\over 3r}+\kappa r$  & ($V^{corr}_{1} in~ ref.[4]$) 
  &($V^{corr}_{1} in~ ref.\cite{lar}$) & & $+V^{corr}_{3}$ \\
 &  &  &  $\beta=0.6$ & $\beta=0.6$ &  & $\beta=0.25$ \\
\hline
$1^1s_0$ & 2978.8$\pm$1.9 & 3074.0 & 2979.3 & 3026.6 & 3010.8 & 2999.5 \\
\hline
$2^1s_0$ & 3594.0$\pm$5.0 & 3662.1 & 3446.3 & 3676.0 & 3625.0 & 3618.4 \\
\hline
$1^3s_1$ & 3096.88$\pm$0.04 & 3074.0 & 3090.6 & 3157.7 & 3095.1 & 3098.8 \\
\hline
$2^3s_1$ & 3686.00$\pm$0.09 & 3662.1 & 3493.1 & 3754.0 & 3674.5 & 3676.7 \\
\hline
$1^3p_0$ & 3415.1$\pm$1.0 &        & 3321.2& 3452.7 & 3440.8 & 3418.0 \\
\hline
$1^3p_1$ & 3510.53$\pm$0.12 &($1P_{c}$)3497.1& 3395.1 & 3533.1 & 3489.3 &
3480.7\\ \hline
$1^3p_2$ & 3556.17$\pm$0.13 &        & 3445.1 & 3605.8 & 3514.4 & 3527.5 \\
\hline
$E_{20}$ & 141.07          &  0     & 123.9  & 153.1  & 73.6   & 109.4 \\
\hline
$E_{21}$ & 45.64           &  0     & 50.0  & 72.7   & 25.1   & 46.7 \\
\hline
$\Delta_{ss}^{(1)}$ & 118.08     &  0    & 111.3   & 131.1  & 84.3  & 99.3 \\
\hline
$\Delta_{ss}^{(2)}$ &  92.0     &  0     & 46.8   & 78.0  & 49.5  & 58.3
\\\hline

\end{tabular}
\end{center}
\end{footnotesize}

\vspace{0.3cm}

\centerline{Table Ib. $b\bar{b}$ system}

\vspace{0.3cm}

\begin{footnotesize}
\begin{center}
\begin{tabular}{|c|c|c|c|c|c|c|}
\hline
 & $~~~~~~~$ & $~~~~~~$ & $V^{Cornell}+V^{corr}_{2}$ &
  $V^{Cornell}+V^{corr}_{2}$ & $~~~~~~~~$ &
    $~~~~~~~~$ \\
 & exp't.  & $V^{Cornell}=$ & $+V^{corr}_{3}+V^{corr}_{1}$ &
  $+V^{corr}_{3}+V^{corr}_{1}$ & $V^{Cornell}+V^{corr}_{2}$ &
   $V^{Cornell}+V^{corr}_{2}$ \\
 &  & ${-4\alpha_s\over 3r}+\kappa r$  & ($V^{corr}_{1} in~ ref.[4]$) 
  &($V^{corr}_{1} in~ ref.\cite{lar}$) & & $+V^{corr}_{3}$ \\
 &  &  &  $\beta=0.6$ & $\beta=0.6$ &  & $\beta=0.25$ \\
\hline
$1^3s_1$ & 9460.37$\pm$0.21 & 9427.0 & 9466.8 & 9503.8 & 9452.8 & 9453.7 \\
\hline
$2^3s_1$ & 100023.30$\pm$0.31 & 10007.0 & 9987.7 & 10073.0 & 10017.0 &
10017.0 \\ \hline
$1^3p_0$ & 9859.8$\pm$1.3   &        & 9843.6 & 9907.6 & 9865.8 & 9861.4 \\
\hline
$1^3p_1$ & 9891.9$\pm$0.7 &($1P_{c}$)9912.8& 9886.1 & 9950.3 & 9902.5 &
9900.8\\ \hline
$1^3p_2$ & 9913.2$\pm$0.6 &        & 9920.5 & 9985.9 & 9929.1 & 9931.7 \\
\hline
$E_{20}$ & 53.4          &  0     & 76.9  & 78.3  & 63.3   & 70.3 \\
\hline
$E_{21}$ & 21.3           &  0     & 34.4  & 35.6   & 26.6   & 30.9 \\\hline

\end{tabular}
\end{center}
\end{footnotesize}

{\footnotesize
In this table, $E_{20}\equiv M_{1^3p_2}-M_{1^3p_0}$,
$E_{21}\equiv M_{1^3p_2}-M_{1^3p_1}$, $\Delta_{ss}^{(1)}\equiv M_{1^3s_1}-
M_{1^1s_0}$ and $\Delta_{ss}^{(2)}\equiv M_{2^3s_1}-M_{2^1s_0}$.}

{\footnotesize
The experimental data are taken from "Partical Physics Booklet", July 1994,
Partical Data Group.}

\vspace{1cm}

The decay width $\Gamma(J/\Psi\rightarrow e^+e^-)$ is also a good test for
various models.
For $\Gamma(e^{+}e^{-})$, the annihilation process is related to
the zero-point wavefunction of $J/\Psi$. Considering the QCD correction, we have
$$\Gamma(e^{+}e^{-})= \Gamma_{0}(e^{+}e^{-})(1-16\alpha_{s}/3\pi)$$
and
$$\Gamma_{0}(e^{+}e^{-})=\frac{16 \pi e^{2}_{Q} \alpha^{2}}{M^{2}}
| \phi(0)|^{2}.$$
where $\phi(0)$ is the zero-point value of the $J/\Psi$ wavefunction.
In this scenario, the wave function $\phi(0)$ undergoes a
modification due to the addition of the non-perturbative QCD terms. With
expression (2) one has $4.85KeV$, while with expression (3), $4.72KeV$ and
the experimental data is $4.69KeV$. Since the zero-point
wavefunction is not sensitive to the new correction, the modified
$\Gamma(e^{+}e^{-})$ does not deviate far from that calculated directly
with the Cornell potential.           \\

\noindent {\bf IV. Discussion and Conclusion}

\vspace{0.5cm}

The potential model is successful in explaining the hadronic spectra and
other properties of heavy quarkonia. First, people believed that quarks
are confined inside hadrons, therefore the potential must include a
confinement term. The simplest form is the linear confinement, $kr$.
Besides, at short-distance, where perturbative QCD works well,
the one-gluon-exchange provides a Coulomb-type potential.
By including these two extreme sides, the Cornell potential, in the form of
$V(r)={-4\alpha_s\over 3r}+\kappa r$ where $\alpha_{s}$ and $\kappa$ are
treated as free parameters, indeed gives reasonable results for
both $c\bar c$ and $b\bar b$ families. However, there must be some
non-perturbative effects which are not included in the linear term of the
Cornell potential and they definitely make substantial contributions to
the evaluation of spectra and other properties.

In the general approaches of Richardson, Fulcher and others, the
universal linear confinement $\kappa r$ were usually kept
unchanged, but the simple propagator of the gluon
${-i\over q^2}(g_{\mu\nu}-{q_{\mu}q_{\nu}\over q^2})$ was modified by
multiplying a $q-$dependent factor $F$ which is model dependent.
Usually, this factor was phenomenologically introduced  based
on some physical arguments  or hints from the lattice calculations or obtained
from the higher order perturbative QCD corrections\cite{gup}.
 
Along the other line \cite{svz}, the QCD sum rules  are also successful in
explaining hadronic effects. It implies that there should be
crossing between two lines. Thus the non-vanishing vacuum condensates which
characterize the non-perturbative effects must be somewhat involved in the
modification factor $F$ and may be the dominant piece or at least an
important one.

In this work, by including the effects of quark and gluon condensates, 
we derive the modified gluon propagator. Namely, a $q-$dependent factor $F$
which is similar to that shown in the recent literature is obtained 
in the framework of QCD. There are no free parameters in the 
derived expressions.

It is important to notice that as pointed out by Shifman, this framework is an
extrapolatation from short distances where perturbative QCD is reliable.
Therefore, one cannot expect that this factor $F$ can 
include as much as a purly phenomenological ansatz. But, it does
shed light on the physical picture and enrich our understanding of
the physical mechanism which binds quarks into hadrons.

For fitting experimental data, it requires that the linear
confinement comes not only from the scalar exchange
but also from the vector exchange. This is consistent with Gupta et al.'s 
model. However, the value of $\beta$ depends on the model. In ref.\cite{gup},
it is about 0.25, but in our case, it must be a value of 0.5$\sim$0.6,
otherwise the result is not meaningful. This indicates that the
vector exchange gives a large contribution to the linear potential.

Our numerical results show that by considering the effects of the quark
and gluon condensates, a more reasonable results, especially a better
fit to the spin splitting $E_{20}, E_{10}$ and $\Delta_{ss}^{(1)}$ can
be obtained. However, $\Delta_{ss}^{(2)}$ does not change in the right
direction. The reason is that we only take the lower dimensional condensates
$<q\bar{q}>$ and $<G^{2}>$ into the consideration. For higher excited
states, interaction range becomes larger, we then have to extrapolate the
scenario to the larger distance by introducing higher dimensional condensates
such as $<q\bar{q}G>$, $<GGG>$ and etc.

There is a discrepancy between refs.[4] and \cite{lar} on the sign
of the coefficient of term $<GG>$. As we pointed above,
maybe it is caused by using different methods or the fixed
point gauge. Unfortunately, the contribution from this term is small
compared to that from $<q\bar q>$, therefore the numerical values calculated
by using the formulae of refs.[4] and \cite{lar} are not
very far apart. We are
going to pursue this problem based on both the first principle and 
phenomenology in our next work.

In our evaluation, the closed-loop corrections are omitted since we only
try to analyze the nonperturbative effect in the present paper. However,
the closed-loop corrections are comparable with those from the quark and
gluon condensates. Therefore, a complete analysis where both the condensate
and loop contributions would be considered will be the aim of our next
work. It will be helpful to clarify that the result as well as the method
in refs. [4] is better or those in ref. [9] is more appropriate.

It should be mentioned that in this work we take the Cornell potential as
the zeroth order approximation. Due to the fact that the wave function at
the origin depends on the model sensitively, it will be interesting to choose
other models as the zeroth order approximation and compare the calculated
results. This will also be done in our next work.

Finally, the nonperturbative effects are considered by using perturbative
treatment. Although it is simple, sometimes it does not work well.
Since our purpose is to illustrate the effects of nonperturbative QCD
effects, it would be not a serious issue. It is also interesting to
employ other ways such as the variational method adopted by Gupta et al
to re-carry out the analysis. It deserves further study.

\vspace{0.7cm}

\noindent {\bf Acknowledgments}

\vspace{0.5cm}

One of the authers (P. Shen) would like to thank Prof. A.W. Thomas for valuable
suggestions and kind hospitality during his stay in the Department of Physics 
and Mathematics, University of Adelaide.

\vspace{0.7cm}

\noindent {\bf Appendix}

\vspace{0.5cm}

By using the method mentioned in the text and ref.[4], the quark-quark
potential is re-derived. The explicit expression of the revised quark-quark
potential can be written as:

\begin{eqnarray*}
U(\vec{p_{1}},\vec{p_{2}},\vec{r})&=&\frac{g^{2}}{4\pi} \frac{(
\lambda_{1}^{a} \cdot \lambda_{2}^{4a})}{4} \Bigg\{
   A_{1} \delta(\vec{r})+
   A_{2} \frac{1}{r}+A_{3}r+A_{4}r^{3}+A_{5\beta}
   \frac{e^{-m_{\beta}r}}{r}\\
&+&\left[ A_{6}\frac{1}{r^{3}}+A_{7}\frac{1}{r}+A_{8}r+A_{8\beta}
   \frac{m_{\beta}^{2}r^{2}+3m_{\beta}r+3}{r^{3}} e^{-m_{\beta}r}
   \right]S_{12}\\
&+&\left[ A_{9}\frac{1}{r^{3}}+A_{10}\frac{1}{r}+A_{11}r+A_{11\beta}
   \frac{m_{\beta}r+1}{r^{3}} e^{-m_{\beta}r}
   \right]\cdot \left[\vec{f_{1}}\cdot(\vec{r}\times\vec{p_{1}})-
   \vec{f_{2}}\cdot(\vec{r}\times\vec{p_{2}}) \right]\\
&+&p-dependent~~~terms \Bigg\}
\end{eqnarray*}

\noindent
with

\begin{eqnarray*}
A_{1}&=&-\pi \left[\frac{1}{2m_{1}^{2}}+\frac{1}{2m_{2}^{2}}-
        \frac{1}{m_{1}m_{2}}+\left( \frac{2}{3m_{1}m_{2}}+\frac{A_{\beta}'}
        {3m_{\beta}^{3}m_{1}m_{2}} \right) (\vec{\sigma_{1}}\cdot
        \vec{\sigma_{2}}) \right]\\
A_{2}&=&1-\frac{A_{\beta}'}{m_{\beta}} \left[\frac{1}{8m_{1}^{2}}+
        \frac{1}{8m_{2}^{2}}-\frac{1}{4m_{1}m_{2}}+\frac{1}{m_{\beta}^{2}}
        +\frac{1}{6m_{1}m_{2}} (\vec{\sigma_{1}}\cdot
        \vec{\sigma_{2}}) \right]\\
A_{3}&=&-\left[\frac{A_{\beta}'}{2m_{\beta}}-B'\left(\frac{1}{16m_{1}^{2}}+
        \frac{1}{16m_{2}^{2}}-\frac{1}{8m_{1}m_{2}}+
        \frac{(\vec{\sigma_{1}}\cdot\vec{\sigma_{2}})}
        {12m_{1}m_{2}} \right)\right]\\
A_{4}&=&\frac{1}{24}B'\\
A_{5\beta}&=&\frac{A_{\beta}'}{m_{\beta}} \left[ \frac{1}{8m_{1}^{2}}+
        \frac{1}{8m_{2}^{2}}-\frac{1}{4m_{1}m_{2}}+ \frac{1}{m_{\beta}^{2}}+
        \frac{(\vec{\sigma_{1}}\cdot\vec{\sigma_{2}})}
        {6m_{1}m_{2}} \right]\\
A_{6}&=&-\left[\frac{1}{4m_{1}m_{2}}-
        \frac{A_{\beta}'}{4m_{1}m_{2}m_{\beta}^{3}} \right] \\
A_{7}&=&-\frac{A_{\beta}'}{24m_{1}m_{2}m_{\beta}}\\
A_{8}&=&-\frac{1}{96m_{1}m_{2}}b'\\
A_{8\beta}&=&-\frac{A_{\beta}'}{12m_{1}m_{2}m_{\beta}^{3}}\\
A_{9}&=&-\left[1-\frac{A_{\beta}'}{m_{\beta}^{3}} \right] \\
A_{10}&=&\frac{A_{\beta}'}{2m_{\beta}} \\
A_{11}&=&\frac{1}{8}B'\\
A_{11\beta}&=&-\frac{A_{\beta}'}{m_{\beta}^{3}},
\end{eqnarray*}

\noindent
where the summation over $\beta$ is implied and the expressions of
$A_{\beta}'$ and $B'$ are given in the text.

\end{document}